\documentclass[journal]{IEEEtran}
\usepackage{cite}
\usepackage{amsmath,amssymb,amsfonts}
\usepackage{algorithmic}
\usepackage{graphicx}
\usepackage{todonotes}
\usepackage{textcomp}
\usepackage{xcolor}
\usepackage{todonotes}
\usepackage{multirow}
\usepackage{bm}
\usepackage{url}
\usepackage{subcaption}
\usepackage{orcidlink}

\IEEEaftertitletext{This work has been submitted to the IEEE for possible publication. Copyright may be transferred without notice, after which this version may no longer be accessible.}
\begin{document}

    \title{Ultra-precise TDoA-based Localization of Frequency Hopping LPWAN Transmitters}
%
%
%
 
\author{Thomas~Maul\,\orcidlink{0009-0008-3295-2993}, 
        Alfred~Mueller\,\orcidlink{0000-0003-3037-4831}, 
        Sebastian~Klob\,\orcidlink{0009-0008-6399-6374},
        and~Joerg~Robert\,\orcidlink{0000-0001-8478-0179},~\IEEEmembership{Member,~IEEE}
\thanks{Thomas Maul is with the Fraunhofer IIS, Nürnberg, Germany, e-mail: thomas.maul@iis.fraunhofer.de}
\thanks{Alfred Mueller was with the Fraunhofer IIS, Fraunhofer Institute for Integrated Circuits IIS, Nuremberg, Germany, email: alfred.mueller@fau.de}
\thanks{Sebastian Klob is with Friedrich-Alexander Universität Erlangen-Nuernberg (FAU), Information Technology (Communication Electronics), Erlangen, Germany, e-mail: sebastian.klob@fau.de}
\thanks{Joerg Robert is with Friedrich-Alexander Universität Erlangen-Nürnberg (FAU), Information Technology (Communication Electronics), Erlangen, and with the Fraunhofer IIS, Nürnberg, Germany, e-mail: joerg.robert@ieee.org}
\thanks{Manuscript received XXXXXX; revised YYYYYYYYYYY.}}

\maketitle

    \begin{abstract}
The Internet of Things (IoT) is a highly emerging market.
It serves as a key enabler for a variety of applications like the digital twin or asset tracking in industrial scenarios.
This often requires the provision of precise position information.
However, systems like Global Navigation Satellite Systems (GNSS) are ruled out due to high energy costs and indoor applications.
A variety of systems is discussed to close this gap.
\\In order to contribute to the investigations of possible gold standards, this paper discusses the localization based on Low Power Wide Area Networks (LPWAN).
Therefore, a concept is presented, based on Time Difference of Arrival (TDoA) measurements within the LPWAN standard ETSI TS 103 357.
This paper addresses two major challenges.
At first, TDoA measurements require highly precise temporal synchronization of the receiving base stations.
Within this work, this issue is solved by exploiting Signals of Opportunity (SoO) as synchronization source, enabling sub-meter synchronization accuracy.
A further issue arises from the Frequency Hopping (FH) waveform of the transmitting endpoints, resulting in a loss of phase information and thus usable localization bandwidth.
A method is introduced to overcome this limitation.
\\This paper states the system concept, proves its functionality in theoretical investigations and simulations.
Finally, real-world measurements verify the functionality and show a 2D localization accuracy of below 10\,m in Line of Sight (LOS) scenarios.
\end{abstract}

    \section{Introduction}
\label{Introduction}
\IEEEPARstart{T}{he} Internet of Things (IoT) is a highly emerging market and forecasts predict an even faster growth in the current decade \cite{IoT_growth}.  
The IoT covers various use cases, from applications in health care over smart cities up to advanced asset management systems for highly autonomous manufacturing. 
A vast majority profits or even preassumes access to position information of the IoT device. 
This has led to a large number of research questions in this field. 
Such approaches are especially necessary since classical methods like Global Navigation Satellite Systems (GNSS)-based systems are infeasible due to energy consumption, costs, and challenging indoor scenarios.
The authors of \cite{IoT_Loc} state an overview of possible localization approaches within the IoT to close the gap. 
However, energy consumption and costs further constrain the localization technology deployed in the field. 
One category of networks, that was especially designed with these constraints in mind, can be found in Low Power Wide Area Networks (LPWAN). 
Prominent representatives of this category are LoRa\footnote{\url{https://www.semtech.com/lora/}, accessed March 2026}, sigfox\footnote{\url{https://sigfox.com/}, accesses March 2026}, and ETSI TS 103 357 \cite[Sec.~6]{ETSI}, known as mioty\footnote{\url{https://mioty-alliance.com/}, accessed March 2026}. These networks allow cost-efficient communication over a large area with battery lifetimes in the range of multiple years. 
An extension for this kind of network concerning localization seems obvious. 

The authors of \cite{LPWAN_Loc} give an overview over different localization approaches within LPWANs. 
The procedures presented can be divided into three areas. 
The most simple approach is the usage of Received Signal Strength Indication (RSSI) values for localization. 
It does neither require any synchronization between the receiving base stations, nor an antenna array. 
However, the authors of \cite{RSSI_0} and \cite{RSSI_2} show that the achievable accuracy is in the range of multiple hundreds of meters, which is not sufficient for many use cases in the context of IoT. 
Furthermore, RSSI-based localization systems are more suited for indoor scenarios, since they profit from complex environments and the resulting multipath propagation. 

Another localization method is based on Angle-of-Arrival (AoA) estimation. This method requires an antenna array at the receiving base stations, allowing an estimation of the incident angle of the received wave. 
The authors of \cite{AoA_0} and \cite{AoA_1} investigate this method for an indoor and outdoor scenario, respectively. 
This method provides better localization results than methods solely relying on RSSI values. However, AoA-based systems require a large number of antennas. 

The last class of localization methods within LPWANs is based on the measurement of runtime differences of the electromagnetic wave from the IoT device to the different receiving base stations. 
These kinds of systems generally deliver the best localization accuracy. 
However, synchronization between the receiving base stations is a necessary prerequisite. 
The authors of \cite{TDoA_1} explain the necessary basics for this Time Difference of Arrival (TDoA) localization within LPWANs. The authors of \cite{TDoA_2} show measurement results from an experimental installation in a city center. 
The accuracy of this localization approach is still in the range of multiple hundreds of meters. However, a non-negligible part of the errors are caused by the synchronization of the base stations using a GNSS-based approach. 

Modern State-of-the-art GNSS-based synchronization modules already deliver good synchronization accuracy in the range of multiple meters, like the ublox NEO-F10T1\footnote{\url{https://www.u-blox.com/en/product/neo-f10t-module}, \newline accessed March 2026}. 
The module's specified precision is 10\,ns, adequate for many localization scenarios but insufficient for sub-meter accuracy.

Another important constraint within TDoA-based localization in LPWAN systems arises from the relatively long transmission durations in the range of seconds when using transmission modes with highest reach.
This requires ultra-precise frequency synchronization to avoid phase drifts during the transmission duration.
The authors of \cite{TDoA_GUC} solved this by direct estimation of the frequency offset.
Further, the authors used an interesting approach by using a transmitter with known position as source of synchronization, operating in the same frequency band.
However, the used principle increases the channel load and causes additional interference within the channel. Further, due to external interference it may completely fail in regions with high channel occupancy in the commonly used license-exempt frequency bands.

To overcome the limitations of the synchronization, we propose a concept for ultra-precise TDoA-based localization of endpoints within LPWANs, that relies on the usage of Signals of Opportunity (SoO) as the source of synchronization.  
This concept was first described in \cite{Troeger}, whereas \cite{Synch} modified this approach for a purely software-based synchronization concept.
The concept achieves a time accuracy in the sub-nanosecond range and a frequency accuracy in the mHz range, which is especially important for the typical high transmission durations within the LPWAN waveforms.

The remainder of this document is structured as follows:
Section \ref{chap:Concept} states a system concept for ultra-precise localization. 
Section \ref{Limits} derives theoretical TDoA estimation limits, whereas Section \ref{Simulation} proves these limits in simulations. 
Section \ref{Measurement} verifies the concept in real-world measurements. 
Finally, section \ref{Conclusion} concludes the document.

    \section{Proposed System Concept}\label{chap:Concept}
This section gives an introduction to the concept of TDoA localization within LPWA networks. It identifies possible hurdles and  states solutions, paving the way for ultra-precise positioning.
\subsection{Multilateration within LPWANs}
\begin{figure}[t]
\centerline{\includegraphics[width=0.5\textwidth]{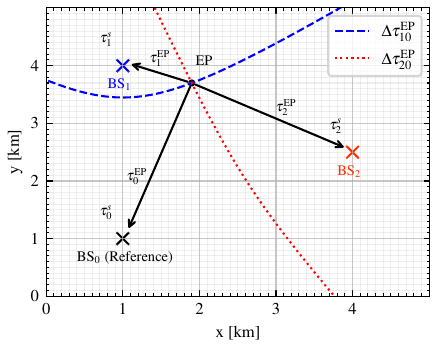}}
\caption{Multilateration scenario for TDoA-based localization of an endpoint (EP) and several base stations (BS) within an LPWA network. The shape of one TDoA value follows a hyperbole \cite{TDoA_2}, whereas the intersection of two hyperboles delivers the position estimate.}
\label{Multilateration}
\end{figure}

We start by considering the classical localization use case, where we want to localize an endpoint in a two dimensional cartesian coordinate system, which is characterized by two unknown variables according to:
\begin{equation}
    \overrightarrow{\text{EP}} = (x_{\text{EP}},y_{\text{EP}}) 
\end{equation}
This scenario is visualized in \mbox{Figure \ref{Multilateration}} with arbitrarily chosen coordinates for the endpoint and a total number of three receiving base stations.
The endpoint radiates a signal that reaches multiple base stations, located at $\overrightarrow{\text{BS}_i} = (x_i, y_i)$, within a certain runtime of the electromagnetic wave $\tau^{\text{EP}}_i$, expressed in seconds.
Instead of measuring the absolute runtime of the electromagnetic wave from the endpoint to the base stations, the difference in the runtime between the endpoint and the different base stations is measured.
This does not require a time synchronization of the endpoint and the respective base stations.
Therefore, the so-called TDoA value $\Delta \tau^{\text{EP}}_{ij}$ can be evaluated by calculating the difference between the runtime of the EP to base station i and base station j \cite{Zhao2020}:
\begin{equation}
    \Delta \tau^{\text{EP}}_{ij} = (\tau^{\text{EP}}_i + \tau_i^S)-(\tau^{\text{EP}}_j+\tau_j^S) 
    \label{eq:tdoa}
\end{equation}
By carefully investigating \eqref{eq:tdoa}, it becomes evident that the difference not only depends on the runtime difference $\tau^{\text{EP}}_i - \tau^{\text{EP}}_j$, but also on two unknowns, namely $\tau_i^S$ and $\tau_j^S$.
These parameters describe the clock offset of the individual base stations to an arbitrarily chosen reference time $t_{\text{ref}}$.
The estimation of this unknown is referred to as synchronization.
The necessity becomes evident when considering the propagation speed of light, denoted as $c$.
An error in the time synchronization of \mbox{3\,ns ($3\cdot 10^{-9}$\,s)} would therefore already result in a position error of \mbox{$3\,\text{ns} \cdot c  \approx 1$\,m}, underlining the need for ultra-precise time synchronization.
After synchronization and computation of the TDoA value, this process has to be computed for a second base station pair in order to achieve an unambiguous position estimate.
This process is reffered to as multilateration process.

Within this article, the main contribution is a precise estimation of the synchronization offset and the estimation of the actual TDoA value.
However, the process of multilateration will be also discussed.

\subsection{Differential TDoA Estimation for Incoherent Transmitters}
In a first step, attention should be drawn to the extraction of the desired runtime values $\tau^{\text{EP}}_i$ from the endpoint to the respective base station.
This requires investigations on the emitted waveform of the endpoint to be localized.
A detailed analysis of the signal structure within the mioty waveform is given in \mbox{Appendix \ref{chap:App_mioty_param}}.
The investigations start by modelling the Frequency Hopping (FH) waveform of the endpoint according to:
\begin{IEEEeqnarray}{rCl}
s_{EP}(nT)
&=& \sum_{b = 0}^{B-1} A \,
     \exp\!\Big(-j(2\pi\epsilon t + \phi_c + \phi_B[b])\Big)
\nonumber \\
&& \cdot
     \exp\!\Bigg(
        j2\pi \sum_{k=0}^{K-1}
        \big(a[k] + 4T f_B[b]\big)
\nonumber \\
&& \qquad\quad \cdot
        g\!\big(t - kT(1+\xi) - \tau\big)
     \Bigg)
\label{eq:ep_dig}
\end{IEEEeqnarray}
Here, $A$ models the signal's amplitude, whereas $\phi_c$ is the global phase offset of the carrier signal and $\epsilon$ is the Carrier Frequency Offset (CFO).
$\bm{\phi}_B$ reflects the vector of random phase offset for each of the B bursts within the telegram, at the respective burst carrier frequency, modeled as vector $\bm{f}_B$.
Further, $k$ is the running index for the K data symbols within a single burst. 
The transmitted payload within a telegram is modeled by the data symbol vector $\bm{a}$, modulated as a Gaussian Minimum Shift Keying (GMSK) waveform with the pulse shape g(t).
Finally, $\tau$ is the delay with respect to an arbitrarily chosen reference time $t_{\text{ref}}$ and $\xi$ models the Sampling Clock Offset (SCO).
Inherently made assumptions on the waveform model are discussed in more detail in the Appendix.
\\Coming back to the localization problem, or more precisely, the extraction of the runtime values from the waveform, the computation of the so-called Cross Correlation Function (CCF) is an often used approach.
Therefore, the endpoint signal $s_{\text{EP}}(t)$ is received at a base station and afterwards correlated with the originally transmitted signal. 
In the digital domain, this can be described according to \cite{Wang2024}:
\begin{equation}
    \hat{R}_{rs}[m] = \frac{1}{N} \sum_{n=0}^{N-1-m} r^*[n]s_{\text{EP}}[n+m]
    \label{eq:R_digital}
\end{equation}
where $m$ states the index within the CCF, while $n$ is the running index of the sample vector with a total of $N$ samples. 
And by further searching for the peak within this CCF, the runtime can be estimated according to:
\begin{equation}
	\hat{\tau}^{\text{EP}} = \underset{m}{\arg\max} (|\hat{R}_{rs}[m]|)
	\label{eq:CCF_max}
\end{equation}
The knowledge of $s_{\text{EP}}(t)$ is a valid assumption, as the payload of the telegram is fully known after passing the decoder.
However, this knowledge does not apply to parameters like the CFO and SCO of the transmitted endpoint signal.
In particular, this does not apply to random phase offsets of the individual bursts.
These offsets occur because the Phase Locked Loop (PLL) is turned off between consecutive bursts to save energy.
After turning the PLL on, the phase locks at a random position.
However, these random offsets severely harm the localization accuracy.
Due to the lost phase information, the localization can simply be executed within a single burst of a telegram.
This limits the accuracy to a fraction of the theoretically achievable when using the whole telegram bandwidth.
\\To overcome this limitation, in the following a concept is applied that is called differential correlation.
The considerations start by a detailed review of the received signal at a base station with index $i$:
\begin{IEEEeqnarray}{rCl}
r_{i}^{\mathrm{EP}}[n]
&=& s_{\mathrm{EP}}\!\Big(
nT\big(1+\xi^S_{i}\big)
+ \tau_{i}^{\mathrm{EP}}
+ \tau^S_{i}
\Big)
\nonumber \\
&& \cdot
\exp\!\Big(
-j\big(
2\pi \epsilon^S_{i} nT
+ \phi^S_{i}
\big)
\Big)
+ n_{i}[n]
\label{eq:RecSignal}
\end{IEEEeqnarray}
where $n_i$ denotes thermal noise added at the receiver.
The parameters $\epsilon^S_i, \xi^S_i, \tau^S_i$, referred to as the synchronization parameters, describe the influences from the erroneous base station oscillator, specifically the CFO $\epsilon^S_i$ and the SCO $\xi^S_i$, within base station $i$.
The delay $\tau_i$, measured at base station $i$ with respect to an arbitrary reference time $t_{\text{ref}}$, is now split into the desired runtime of the electromagnetic wave $\tau_{i}^{\text{EP}}$ and the temporal offset of the local oscillator $\tau_{i}^S$.
\\The concept of differential correlation now stipulates that the signals received from two base stations are correlated against each other.
Analogous to \eqref{eq:R_digital}, the CCF between these base stations can be calculated according to:
\begin{IEEEeqnarray}{rCl}
\hat{R}^{\mathrm{EP}}_{r_0 r_1}[m]
&=& \frac{1}{N}
\sum_{n = 0}^{N-1-m}
(r_{0}^{\mathrm{EP}}[n])^*\,
r_{1}^{\mathrm{EP}}[n+m]
\nonumber \\[4pt]
&=& \frac{1}{N}
\sum_{n = 0}^{N-1-m}
s_{\mathrm{EP}}^*\!\Big(
nT(1+\xi^S_{0})
+ \tau_{0}^{\mathrm{EP}}
+ \tau^S_{0}
\Big)
\nonumber \\[4pt]
&& \cdot
s_{\mathrm{EP}}\!\Big(
nT(1+\xi^S_{1})
+ \tau_{1}^{\mathrm{EP}}
+ \tau^S_{1}
+ mT
\Big)
\nonumber \\[4pt]
&& \cdot
\exp\!\Big(
-j2\pi nT(\epsilon^S_{1}-\epsilon^S_{0})
\Big)
\nonumber \\[4pt]
&& +\, n^{\mathrm{EP}}_R[n]
\label{eq:ccf_diff}
\end{IEEEeqnarray}
Interestingly, the CCF now solely consists of $s_{\text{EP}}[n]$ and $s^*_{\text{EP}}[n]$, or more precisely a product of both.
Instead of correlating the received signal with the reconstructed prototype function, the differential correlation can be executed blindly without prior reconstruction of the transmitted signal.
All interfering parameters from the waveform like the random phase offsets of the individual bursts and the effects from the erroneous oscillator of the endpoint do not have to be estimated, as they cancel out during the calculation of the differential CCF.
The CCF now depends on the propagation delays of the electromagnetic waveform from the endpoint to the respective base station.
The difference is the wanted TDoA value $\Delta \tau^{\text{EP}}_{10}$:
\begin{equation}
    \Delta \tau^{\text{EP}}_{10} = \tau_{0}^{\text{EP}} - \tau_{1}^{\text{EP}}
    \label{eq:delta_tau}
\end{equation}
And by searching for the peak within the CCF, the TDoA value can be estimated according to \eqref{eq:CCF_max}:
\begin{equation}
    \Delta\hat{\tau}^{\text{EP}}_{10} = \underset{m}{\arg\max} \Big\{ \Big|\hat{R}^{\text{EP}}_{r_0r_1}[m] \Big| \Big\}
\label{eq:tdoa_extraction}
\end{equation}
This is a very interesting finding as it unveils that the localization can be executed for various FH waveforms, not only for the initially considered mioty waveform.
The stated solution can be transferred to a variety of different frequency-hopping systems, e.g. Bluetooth.
Another advantage is that the transmitting source does not need to be cooperative, as modifications within the waveform, like random phase offsets or encrypted data encoding, are not of relevance due to the differential correlation.
This puts especially millitary use cases in the focus, where one might be interested in the localization of non cooperative communication systems, e.g. the localization of drones to name just one.
Future work should therefore include the transfer of the stated approach to different systems and use cases.

The following considerations, however, focus on the already introduced LPWAN standard mioty.
By taking a closer look at \eqref{eq:ccf_diff}, it becomes evident that the CCF still depends on the parameters caused by erroneous oscillators and a combined noise process $n^{\text{EP}}_R[n]$ of both thermal noise components from both base stations.
To obtain an exact estimate of $\Delta \hat{\tau}^{\text{EP}}_{10}$, a set of parameters must be estimated first.
These unwanted parameters are the temporal offsets of the base station oscillators to the arbitrarily choosable reference time $t_{\text{ref}}$, denoted as $\tau^S_{0}$ and $\tau^S_{1}$.
Further, the CCF depends on the CFO difference $\epsilon^S_0 - \epsilon^S_1$ and the SCO $\xi^S_{0}$, respectively $\xi^S_{1}$. 
\\The estimation of these parameters is referred to as synchronization.
In the following section a synchronization approach is presented, that outperforms the accuracy of currently available solutions relying on GNSS.
\subsection{Extraction of Synchronization Parameters}
The synchronization concept used within this article is presented in more detail in \cite{Synch}.
The main idea behind the stated concept is the usage of uncooperative broadcast signals for time and frequency synchronization.
These signals may be audio broadcast ("DAB": Digital Audio Broadcast\cite{DAB}) or television broadcast ("DVB-T": Digital Video Broadcast - Terrestrial\cite{DVBT}).
However, the article highlights that almost any signal might serve as SoO.
Potential further candidates are cellular systems like 5G, or 6G in the future.
\\The core idea of the concept is the reception of the SoO in a second channel in parallel to the localization waveform (in our case the mioty signal).
If the dual-channel frontend has the so-called LO-sharing capability, where one oscillator feeds both reception channels, the synchronization parameters can be converted between both channels.
This LO-sharing capability is present in many state of the art Software Defined Radio Frontends (SDR), like the SDRPlay RSPduo\footnote{\url{https://www.sdrplay.com/rspduo/}, accessed March 2026} or the Ettus TwinRX daughterboard\footnote{\url{https://files.ettus.com/manual/page_twinrx.html}, accessed March 2026}.
The article \cite{Synch} proposes a purely software-based algorithm to extract the desired synchronization parameter from the IQ samples of the SoO.
This is done by computing the differential CCF between the base stations to be synchronized.
This blind correlation approach does not require any decoding of the data symbols from the SoO, rendering this approach quite flexible.
After an estimation of the synchronization parameters based on the SoO waveform, they can be converted to the second channel.
If the same sampling rate is used in both reception channels, the SCO can be converted directly:
\begin{equation} 
    \Delta \hat{\xi}_{10}^{S} = \hat{\xi}_{1}^{S} - \hat{\xi}_{0}^{S}= \Delta \hat{\xi}_{10}^{\text{SoO}}
\label{eq:xi_conversion}
\end{equation} 
As the timing offset (TO) and the SCO originate from the same erroneous sampling interval, the relation for the TO can be given straight forward:
\begin{equation} 
    \Delta \hat{\tau}_{10}^{S} = \hat{\tau}_{1}^{S} -\hat{\tau}_{0}^{S} = \Delta \hat{\tau}_{10}^{\text{SoO}}
\label{eq:tau_conversion}
\end{equation} 
For the conversion of the CFO, the different carrier frequencies have to be taken into account.
Although the relative frequency error of the oscillator is the same in both channels, the absolute frequency error scales with the set carrier frequency in the respective channel. 
A higher carrier frequency therefore leads to a higher absolute frequency deviation as follows:
\begin{equation}
	\Delta \hat{\epsilon}_{10}^{\text{S}} = \hat{\epsilon}_{1}^{\text{S}} -\hat{\epsilon}_{0}^{\text{S}} = \Delta \hat{\epsilon}^{\text{SoO}} \cdot \frac{f_{\text{EP}}}{f_{\text{SoO}}},
\label{eq:epsilon_conversion}
\end{equation}
where $f_{\text{EP}}$ and $f_{\text{SoO}}$ denote the carrier frequencies of the mioty waveform and the SoO.
\\After carefully reviewing \eqref{eq:ccf_diff} again, the TDoA estimation can be directly executed after compensating the parameters found in \eqref{eq:xi_conversion} to \eqref{eq:epsilon_conversion}.
While the compensation of the CFO is straightforward by multiplying an exponential function to the raw IQ samples, a more sophisticated method is required to compensate for the SCO.
The compensation requires fractional resampling.
In the case at hand, this is achieved by using a Farrow structure, as introduced in \cite{Farrow}.
After applying the Farrow resampling, the CCF can be calculated and the TDoA value can be estimated according to \eqref{eq:tdoa_extraction}.

    \section{Theoretical Limits of TDoA Estimation} \label{Limits}
After proposing a concept for the TDoA estimation of incoherent endpoints in LPWAN  in the previous section, the theoretically achievable accuracy has to be investigated in detail.
The investigations are divided into two parts.
At first, the so-called Modified Cramer Rao Lower Bound (MCRB) is evaluated to achieve an estimate on the lowest achievable estimation variance.
In a second step, the shape of the CCF will be examined in detail, since the estimation of the TDoA value directly depends on it.
\subsection{Lowest Estimation Variance according to MCRB}
\begin{figure}[t]
\centerline{\includegraphics[width=0.5\textwidth]{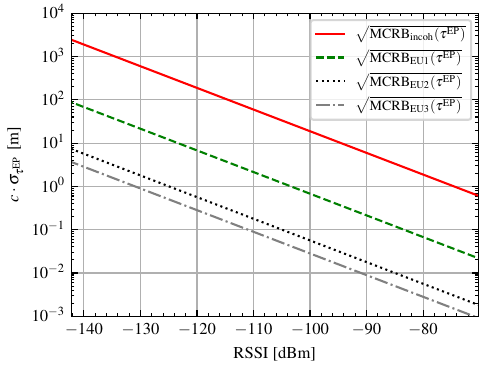}}
\caption{MCRB for the estimation of the parameter $\tau^{\text{EP}}$ for the incoherent estimate according to \eqref{eq:MCRB_TAU_INCOH}, and for the different bandwidth modes according to \eqref{eq:MCRB_TAU} as a function of the RSSI.}
\label{fig:mcrb_tdoa}
\end{figure}

The Cramer Rao Bound (CRB) is an often-used metric in estimation theory.
It states the lowest achievable variance for an unbiased estimator.
However, as \cite{MCRB} shows, the calculation of this limit is often not feasible due to analytically unsolvable integrals resulting from an often unknown probability density function (pdf).
Thus, the authors of \cite{MCRB} stated the Modified Cramer Rao Bound (MCRB), which solves this issue by using a more easily available composite pdf.
Despite the advantage that the MCRB is far more easily accessible than the CRB, it tends to be less tight, or in other words, too optimistic.
Assuming the estimation of an unbiased parameter $\lambda$, this can be expressed as:
\begin{equation}
    \sigma^2(\hat{\lambda} - \lambda) \geq \text{CRLB}(\lambda) \geq \text{MCRB}(\lambda)
\end{equation}
Despite this disadvantage, the MCRB generally serves as a good starting point to assess the theoretically achievable lowest variance.
Transferred to our localization problem, the MCRB can be derived for the estimation of the desired runtime values $\tau^{\text{EP}}$ from the mioty waveform.
A detailed derivation is given in \mbox{Appendix \ref{chap:App_MCRB_Derivation}}.
It follows:
\begin{equation}
    \text{MCRB}(\tau^{\text{EP}}) = \frac{2T^2}{\pi^2 L \, B(16 T^2 \sigma_{f_B}^2 + 1)}\frac{N_0}{E_S},
\label{eq:MCRB_TAU}
\end{equation}
where $T$ is the symbol duration of the mioty waveform, $L$ is the total duration of one burst in symbols, and $N_0/E_S$ is the inverse of the symbol to noise ratio.
Nevertheless, the most interesting influencing parameter for the MCRB can be found in $\sigma_{f_B}^2$.
It states the variance of all carrier frequencies used within the telegram.
It also indicates that an increase in variance within the carrier frequencies results in a lower estimation variance, and thus a lower localization accuracy.
This is also very intuitive, as it indicates that an increased bandwidth leads to a better localization accuracy.
In contrast, \eqref{eq:MCRB_TAU} allows us to estimate the achievable accuracy for the incoherent estimate where we can simply perform the localization by using single bursts for the TDoA estimation.
This virtually allows the usage of a single bursts bandwidth.
By setting $\sigma_{f_B}^2 = 0$, it follows:
\begin{equation}
    \text{MCRB}_{\text{incoh}}(\tau^{\text{EP}}) = \frac{2T^2}{\pi^2 L \,B}\frac{N_0}{E_S}
\label{eq:MCRB_TAU_INCOH} 
\end{equation}
In a next step, the achievable accuracies shall be examined under investigation of the different influencing parameters from \eqref{eq:MCRB_TAU}.
However, most parameters are already fixed by the standardized waveform according to \cite{ETSI}.
Therefore, \mbox{Appendix \ref{chap:App_mioty_param}} gives an overview of the standardized mioty waveform and important parameters for localization.
The most important parameter can be found in the so-called bandwidth modes EU1/EU2/EU3, differing in the total amount of the used hopping bandwidth.
\\However, one parameter is not fixed, but strongly influences the achievable accuracy, namely the signal strength at the receiving base station.
Therefore, \mbox{Figure \ref{fig:mcrb_tdoa}} shows the MCRB according to \eqref{eq:MCRB_TAU} and \eqref{eq:MCRB_TAU_INCOH} for the different bandwidth modes (EU1, EU2, EU3) as a function of the RSSI.
For this purpose, the $E_S/N_0$ has been transformed to the RSSI by taking into account the symbol rate. Further, the minimum RSSI within the graphic has been chosen to match a theoretical Packet Error Rate (PER) of 10\,\%, virtually describing the decoding limit.
Below this threshold, a localization is of little use.
\\\mbox{Figure \ref{fig:mcrb_tdoa}} underlines the inferior performance of the incoherent estimate.
At the decoding limit, the theoretically achievable accuracy ranges at multiple kilometers, rendering localization infeasible.
Only for high reception levels, localization becomes possible. 
Considering the narrow bandwidth mode according to EU1, the theoretically achievable performance drastically increases.
However, localization at the decoder limit is still barely possible.
Sub-meter accuracy can only be achieved for \mbox{RSSI $>$ -104\,dBm}.
The most promising candidate, currently specified in \cite{ETSI}, is the wideband pattern EU2.
It allows reasonable accuracy even at the decoder limit, sub-meter localization is possible even for low reception levels of \mbox{RSSI $>$ -125\,dBm}.
\\Despite these already promising findings, the discussion from \mbox{Appendix \ref{chap:App_mioty_param}} highlighted that the bandwidth can be doubled with minor modifications to further enhance the expected accuracy.
The newly postulated ultra wideband pattern, assigned to bandwidth mode EU3, is fully compatible to the currently deployed decoder infrastructure.
It allows sub-meter estimation accuracy down to \mbox{RSSI $>$ -130\,dBm}.
\\However, one thing was not taken into account during these considerations.
The derivation and evaluation of the MCRB was done solely for the estimation of the propagation time from the node to one base station ($\tau^{\text{EP}}$).
This is only partially correct, as TDoA localization requires the estimation of the runtime difference between pairs of base stations, as described in \eqref{eq:tdoa}.
This becomes especially evident when considering the differential correlation approach stated earlier in this article.
This fact must be kept in mind in the following considerations, even though the derived MCRB promises sub-meter localization accuracy near the decoding limit.
\subsection{Shape of the Autocorrelation Function}
\begin{figure}[t]
\centerline{\includegraphics[width=0.5\textwidth]{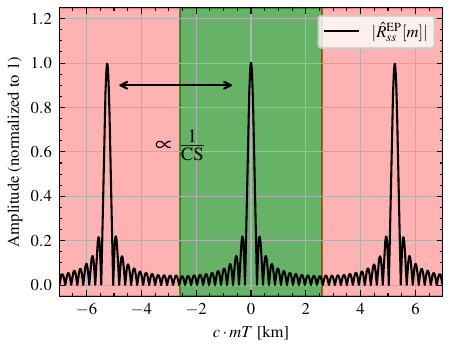}}
\caption{ACF according to \eqref{eq:R_digital} of a mioty telegram in bandwidth mode EU3. The distance between the peaks is proportional to the inverse of the carrier spacing. The green marking represents the area of the correct chosen hypothesis, while the red marking shows the areas where a wrong hypothesis is chosen.}
\label{fig:ACF}
\end{figure}

After stating the lower limit for the estimation variance of the TDoA values, further attention should be paid to the shape of the CCF.
This is especially important as the TDoA values itself are directly extracted from this CCF by performing a peak search according to \eqref{eq:tdoa_extraction}.
Thus, the shape of the CCF directly influences this accuracy.
In the following, a qualitative analysis of the shape of this CCF is done.
Especially, the effect arising from the pattern structure within the mioty waveform is to be investigated.
\\To this end, a telegram is simulated with an empty payload described by the newly postulated ultra wideband pattern EU3.
The ACF, denoted as $\hat{R}_{ss}^{\text{EP}}$[m], is calculated according to \eqref{eq:R_digital} and visualized in \mbox{Figure \ref{fig:ACF}}.
\\The findings of \mbox{Figure \ref{fig:ACF}} unveils interesting effects.
At first, the ACF shows not only a single peak but multiple peaks with even spacing.
By detailed investigation, the distance between these peaks can be traced back to a constant shift being proportional to the inverse of the carrier spacing.
This is caused by the sparse occupation within the frequency domain (see Figure \ref{fig:mioty_pattern}). 
Even though the telegram has a wide bandwidth of 1.31\,MHz in the EU3 mode, the whole spectrum consists only of 24 narrow bursts (2.38\,kHz).
However, this finding also hints that the ambiguities vary for the different bandwidth modes, as their CS parameter also changes.
Due to this link, the problem of ambiguities will be most pronounced for the broadband pattern EU3.
Coming back to the localization problem, this issue is critical as it increases the likelihood of choosing the wrong peak and thus, may lead to localization errors in the range of multiple kilometers.
However, this can be prevented by restricting the search area within the CCF to the unambigious region (represented in green within \mbox{Figure \ref{fig:ACF}}).
Expressed as a formula, the following applies to the possible TDoA hypotheses:
\begin{equation}
    \Delta \tau_{\text{hypo}} [\text{m}] \in \Big\{-\frac{c}{2\cdot CS}, \frac{c}{2\cdot CS}\Big\}
\end{equation}
For the ultra wideband pattern EU3, it follows \mbox{$\Delta \tau_{\text{hypo}} \in \big\{-2.62\,\text{km}, 2.62\,\text{km}\big\}$}.
For wider installations in bigger cities, this might already be critical as the distance between base stations, and thus, the maximum TDoA value, may exceed this unambigious area.
In the case at hand, the further considerations are restricted to the search area as stated above, virtually ruling out any ambiguities.
\\However, future work might consider methods to resolve these ambiguities to allow for bigger installations.
Possible approaches include sensor data fusion with RSSI information or AoA measurements.

    \section{Simulative Verification of the Proposed Concept}
\label{Simulation}
This section aims to provide first simulation results for further investigations on the achievable localization accuracy under real influencing parameters like AWGN and residual frequency offsets.
\subsection{Data Processing Pipeline}
\begin{figure}[t]
\centerline{\includegraphics[width=0.5\textwidth]{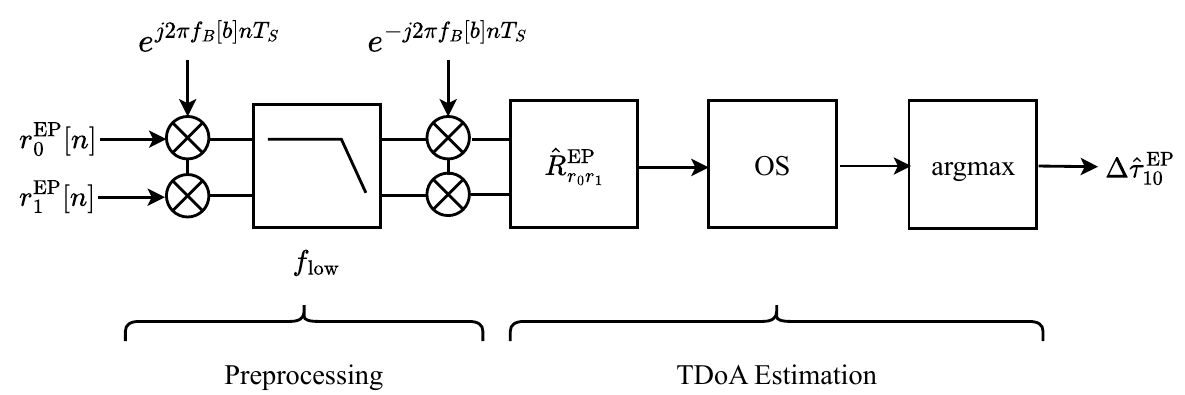}}
\caption{Data processing pipeline from the raw IQ samples of two base stations to the estimated TDoA value.}
\label{fig:preprocessing}
\end{figure}

Prior to the simulations, this section aims to give a detailed introduction to the data processing pipeline.
This concerns the single processing steps from the raw IQ samples to the estimated TDoA value.
A visualization of this pipeline is given in \mbox{Figure \ref{fig:preprocessing}}.
\\The pipeline starts with the fully synchronized IQ samples from two base stations ($r^{\text{EP}}_0[n]$ and $r^{\text{EP}}_1[n]$).
Next, a preprocessing step is executed.
Therefore, each burst is mixed to \mbox{$f=0$\,Hz} and a lowpass filter with a cutoff frequency five times the burst bandwidth is deployed.
This step can be explained with the typical environment LPWAN systems are operating in.
Usually, the license exempt 868\,MHz frequency band is used by a variety of communication systems, potentially causing severe interference on the localization accuracy.
By employing this lowpass filter, interference outside the single bursts can be discarded.
After filtering, the bursts are shifted back to their original position.
\\In the next step, the actual TDoA estimation is executed.
The CCF is estimated according to \eqref{eq:ccf_diff}.
Before searching for the peak within the CCF according to \eqref{eq:tdoa_extraction} to determine the TDoA value, oversampling with a factor of 256 is applied to enhance the resolution within the CCF.
The pipeline terminates with the estimated TDoA value $\Delta\hat{\tau}^{\text{EP}}_{10}$.
\subsection{Estimation Accuracy in AWGN Channel}
\begin{figure}[t]
\centerline{\includegraphics[width=0.5\textwidth]{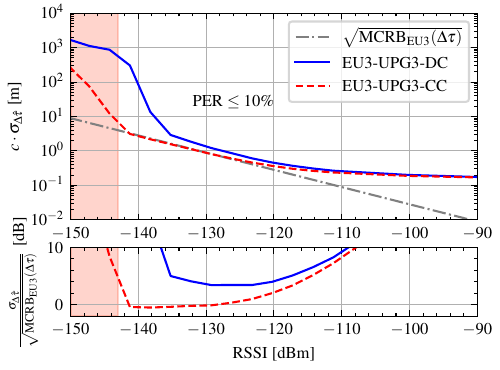}}
\caption{Upper: Simulation of the modified wideband mode (EU3) with low delay pattern (UPG3) for coherent correlation (CC) and differential correlation (DC) as a function of the RSSI in comparison to the MCRB. The red area represents the range of RSSI values resulting in a PER greater 10\,\%.\\ Lower: Distance of the respective simulation to the MCRB in dB ($20\cdot  \text{log}_{10}(.)$). N = 10.000 for each RSSI value.}
\label{fig:sim_awgn}
\end{figure}

In a first simulation, the proposed concept shall be evaluated under pure AWGN influence.
In order to assess the achieved accuracy, a comparison to the MCRB is investigated.
Therefore, the simulation is conducted for different RSSI values. 
A further investigation concerns the differential correlation approach.
As shown in \mbox{Chapter \ref{chap:Concept}}, instead of correlating the received telegram at a base station with the noise-free reconstructed prototype function, it is correlated with the noisy signal of another base station.
This simulation therefore aims to quantify the loss caused by an increased noise power within the correlation process.
\mbox{Figure \ref{fig:sim_awgn}} shows the measured standard deviations for different RSSI values for a correlation of the receive signal with a noise-free prototype function (EU3-UPG3-CC) as well as the correlation of two noisy receive signals (EU3-UPG3-DC).
As \mbox{Chapter \ref{chap:Concept}} already hinted a superior performance of the EU3 bandwidth mode, it is used during this simulation.
Furthermore, the simulation is solely executed for the low latency pattern UPG3.
However, under pure AWGN influence it is expected that the performance of the UPG1 and UPG3 patterns is identical.
This becomes evident, when considering that the total energy within the UPG1/3 patterns is identical and thus, the derived MCRB is identical.
However, this is only valid if no time-dependent effects are investigated.

Interesting observations can be obtained from \mbox{Figure \ref{fig:sim_awgn}}.
At first, the correlation of the received signal with the prototype function (coherent correlation) is considered.
Only very close to the decoding limit, the standard deviation drastically increases.
However, for \mbox{RSSI $>$ -140\,dBm} the measured standard deviation is exactly on the simulated MCRB.
This proves the correctness of the MCRB as well as its tightness compared to the CRB.
For further increasing RSSI values, the simulated curve is flattening out, most likely due to numerical inaccuracies.
However, this flattening happens at approximately 0.2\,m, and is therefore not expected to negatively influence the accuracy.
\\Another observation within \mbox{Figure \ref{fig:sim_awgn}} concerns the curve for the differential correlation.
In contrast to the coherent correlation, the performance detoriates for low RSSI.
Especially near the decoding limit, the measured standard deviation drastically increases.
Only for \mbox{RSSI $>$ -135\,dBm}, the distance to the MCRB shrinks to a minimum of 3\,dB before the inaccuracies cause a flattening for further increasing RSSI values.
The loss of 3\,dB is a very intuitive finding because the correlation of two signals with even noise power thus leads to a doubling of the noise and thus, a loss of 3\,dB in the standard deviation as described by the MCRB.
However, this does not explain why the loss exceeds this value for decreasing RSSI values, potentially indicating a non-linear behavior within this process.
Future work should conduct further investigations on this effect.
This is especially the case as the investigations above solely assumed a correlation of signals with identical noise levels, which is not realistic in real-world scenarios.
In reality, one signal is likely to be stronger than the other.
Most likely, the achievable accuracy will therefore be dominated by the weaker signal.
\subsection{Estimation Accuracy with Oscillator Impairments}
\begin{figure}[t]
\centerline{\includegraphics[width=0.5\textwidth]{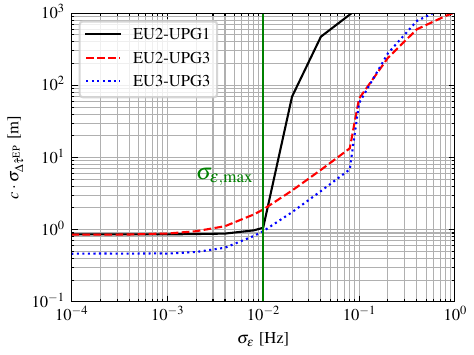}}
\caption{Simulated TDoA estimation accuracy as a function of the frequency standard deviation for the wideband profile (EU2 - pattern group UPG1 and UPG3) and for the modified ultra-wideband mode (EU3-UPG3), simulated at RSSI =  -120\,dBm. N = 10.000 simulations for each standard deviation.}
\label{fig:sim_oscillator}
\end{figure}

Even though the simulations already promise good accuracy in the sub-meter range, the assumption of pure AWGN noise is too simplistic.
Thus, further simulations are carried out with more realistic influencing parameters, namely oscillator impairments.
As \mbox{Chapter \ref{chap:Concept}} already highlighted, errors resulting from noisy oscillators are the main error source within the CCF beside the already considered AWGN.
The following simulations assume that the frequency synchronization between two base stations is no longer perfectly possible.
Therefore, different standard deviations of the frequency synchronization mismatch ($\sigma_{\epsilon}$) are simulated, ranging from $10^{-4}$\,Hz to 1\,Hz.
As this effect is time-dependent, the simulation is carried out for the different bandwidth modes and UPG modes.
Finally, \mbox{Figure \ref{fig:sim_oscillator}} shows the measured standard deviation of the TDoA values for different standard deviations of the residual frequency offset for the different modes with a fixed \mbox{RSSI = -120\,dBm}.
\\From \mbox{Figure \ref{fig:sim_oscillator}} it becomes evident, that the different bandwidth modes and UPG modes no longer behave identically.
At first, the standard bandwidth mode EU2 is considered.
Only for $\sigma_{\epsilon} < 10^{-3}$\,Hz, the standard transmission duration according to UPG1 and the low latency transmission (UPG3) behave identical.
In this area, the frequency synchronization mismatch has minor influence on the localization accuracy.
However, after $\sigma_{\epsilon}$ passes $10^{-2}$\,Hz, the standard deviation for the UPG1 pattern drastically increases.
In contrast, the UPG3 profile longer delivers reasonable TDoA estimation accuracy.

This effect can be directly explained by the transmission duration itself.
Frequency offsets cause a time dependent phase shift.
The longer the telegram, the stronger is the drift between the first and last burst within the telegram.
These drifts cause a deterioration of the CCF.
Thus, shorter transmission durations are less subsceptible against frequency offsets than longer transmission durations.
Similiar observations can be made for the bandwidth mode EU3.
For increasing $\sigma_{\epsilon}$, the standard deviation of the TDoA estimate increas as well.
For $\sigma_{\epsilon} \approx 10^{-1}$\,Hz, the gain of the increased bandwidth in comparison to EU2 tends to 0.
At this point, the oscillator impairments  are the main contributor to the localization error.
\\In summary, it can be concluded that the long transmission duration of mioty telegrams poses strict requirements on the frequency synchronization accuracy.
However, by ensuring $\sigma_{\epsilon} < 10^{-2}$\,Hz, a standard deviation of the TDoA values in the sub-meter range can be expected.

    \section{Initial Measurement Results}
\label{Measurement}
This section provides initial measurement results in real-world environments.
This includes cable-based measurements in a first step, followed by measurements in an urban city area.
\\The architecture of the testbed including the software and hardware components is presented in detail in \cite{Testbed}.
The article further presents initial measurement results for the synchronization concept.
\subsection{Cable-based Measurements in the Laboratory}
\begin{figure}[t]
\centerline{\includegraphics[width=0.5\textwidth]{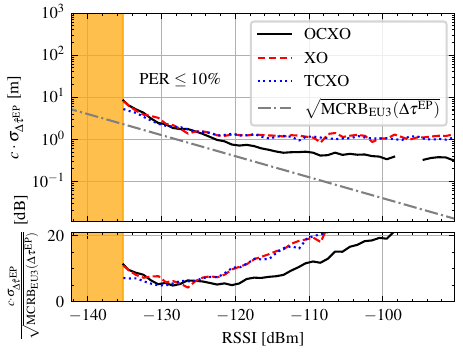}}
\caption{Comparison of the measured standard deviation of the TDoA values $c\cdot \sigma_{\Delta \hat{\tau}^{\text{EP}}}$ for different oscillator types as a function of the RSSI for the bandwidth mode EU3. The lower plots visualize the distance of the measured standard deviation from the MCRB, given in dB.}
\label{fig:meas_lab}
\end{figure}

In a first step, the proposed system concept including the synchronization and the TDoA estimation shall be evaluated in a cable-based environment.
The cable-based measurements allow to rule out the influence of multipath propagation.
Further, the RSSI can be easily controlled.
In the following, this is used to achieve a comparison of the measured standard deviation with the theoretical findings from \mbox{Figure \ref{fig:mcrb_tdoa}}.
Therefore, an endpoint is programmed to transmit telegrams in bandwidth mode EU3 and low latency pattern UPG3 as it is expected to deliver the best robustness against residual frequency offsets.
The signal from the node is split and fed into the inputs of two independent SDR receivers.  
The SDRs are the commercial off-the-shelf (COTS) and low-cost (250\$) SDRPlay RSPduo\footnote{https://www.sdrplay.com/rspduo/, accessed March 2026}.
Furthermore, a step attenuator is placed in the chain and is programmed to increase the attenuation until the telegrams can no longer be received.
As synchronization waveform, a DAB signal (channel 5C, $f_c$ = 178.352\,MHz) is received with a single antenna and fed into both SDRs.
\\Another investigation concerns the oscillator used during the measurements.
As \cite{Synch} already highlighted, the oscillator plays a major role for the achievable synchronization accuracy.
This is especially critical as the simulations conducted above already showed the strict requirements on the frequency synchronization accuracy to achieve a low standard deviation within the TDoA measurements.
To this end, three different oscillator types are investigated during the measurements.
The different oscillator types and exact models are extensively described in \cite{Synch}.
\\Finally, \mbox{Figure \ref{fig:meas_lab}} shows the standard deviations for the TDoA estimation as a function of the RSSI for the different oscillator types (XO, TCXO, OCXO).
In comparison to the measurements, the MCRB for the bandwidth mode EU3 from \mbox{Figure \ref{fig:mcrb_tdoa}} is visualized.
The orange area depicts the RSSI range with a PER of 10\% and higher. 
\\Interesting conclusions can be drawn from the observations in \mbox{Figure \ref{fig:meas_lab}}.
At first, the RSSI range near the decoding limit is considered.
Here, the different oscillator types behave similiar.
This suggests that AWGN noise dominates as an influence here.
Nevertheless, the estimation accuracy of the TDoA values is reasonably good with a maximum of approximately 10\,m.
For increasing RSSI, the standard deviation decreases for all oscillator types.
However, for \mbox{RSSI $>$ - 124\,dBm}, an interesting effect occurs.
Only the OCXO achieves lower standard deviations, while the curves for the XO and the TCXO start to flatten out.
At this point, the main influencing factor is no longer the pure AWGN but the frequency instability of the oscillators.
Only the very frequency-stable OCXO allows to meet the requirements to achieve a sub-meter standard deviation of 0.3\,m.
\\Another interesting observation from \mbox{Figure \ref{fig:meas_lab}} concerns the comparison of the measured TDoA values with the MCRB. 
The MCRB is reached with a minimum distance of \mbox{4.9\,dB for RSSI = -130\,dBm}.
This proves that the stated TDoA estimation method is operating near the optimum.
\subsection{Measurements in Real-World Environment}
\begin{figure}[t]
\centerline{\includegraphics[width=0.48\textwidth]{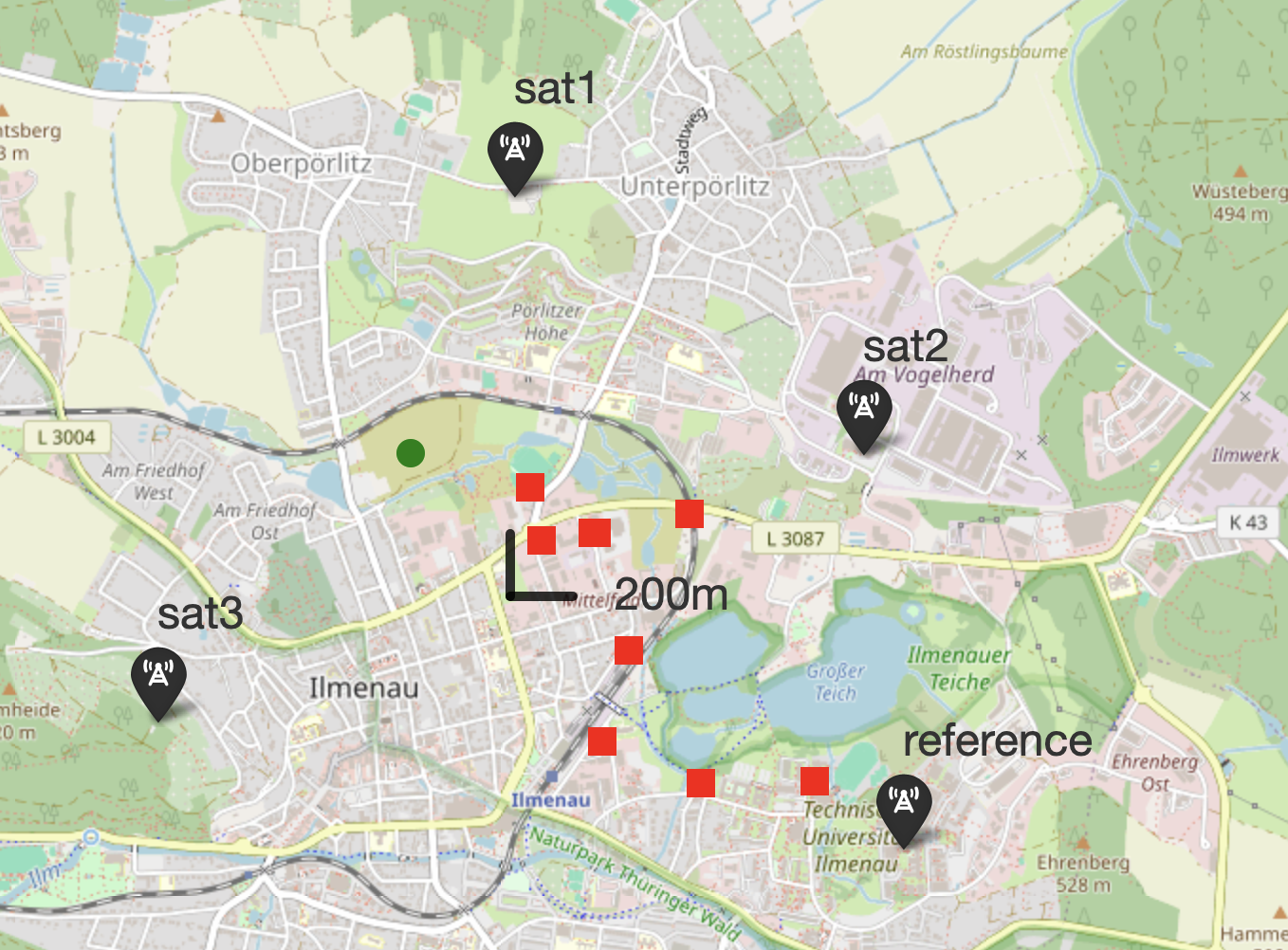}}
\caption{Constructed testbed with four base stations in the city center of Ilmenau (Thuringia, Germany). The testbed covers the whole city center and has a size of approximately $3\,\text{km} \,\text{x}\, 3\,\text{km} \,(9\,\text{km}^2)$. Tested endpoint positions are visualized in green (LOS) and red (NLOS). The transmit power of the endpoint was set to 13\,dBm (20\,mW). @OSM-contributors.}
\label{fig:testbed}
\end{figure}

\begin{figure}[t]
\centering
\begin{subfigure}{0.48\textwidth}
    \centering
    \includegraphics[width=\linewidth]{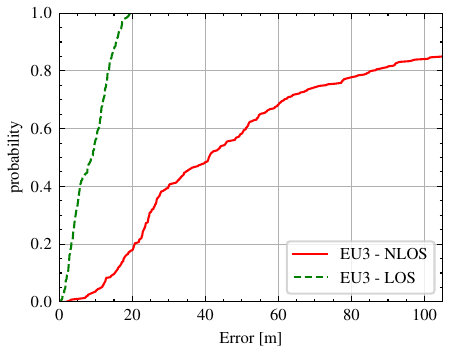}
    \caption{Comparison of the CDF for the endpoint under LOS influence and the endpoints under NLOS influence. The CDF for the NLOS case is zoomed to the first 150\,m.}
    \label{fig:meas_ilm_nlos_zoomed}
\end{subfigure}
\hfill
\begin{subfigure}{0.48\textwidth}
    \centering
    \includegraphics[width=\linewidth]{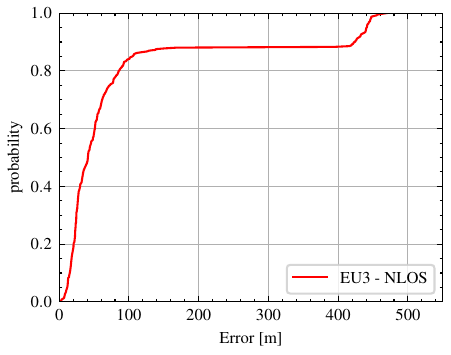}
    \caption{Complete CDF for the endpoints under NLOS condition.}
    \label{fig:meas_ilm_nlos}
\end{subfigure}
\caption{CDF of the 2D localization error for the evaluated endpoint positions as given in \mbox{Figure \ref{fig:testbed}} under LOS and NLOS influence.}
\label{fig:meas_ilm_nlos_combined}
\end{figure}

After proving the functionality of the proposed system concept for precise TDoA estimation in a cable-based environment, the functionality shall be proven in a wide are installation in the next step.
To this end, a testbed was installed in a city center in Ilmenau (Thuringia, Germany).
The testbed consists of four base stations and covers an area of approximately $3\,\text{km} \,\text{x}\, 3\,\text{km} \,(9\,\text{km}^2)$.
An overview of this installation is given in \mbox{Figure \ref{fig:testbed}}.
\\In order to obtain an estimate of the achievable localization accuracy, the localization is executed for nodes positioned at several locations.
In order to assess the influence of multipath propagation on the localization accuracy, two different scenarios are investigated.
This includes a position with Line of Sight (LOS) to all base stations, as well as multiple positions with at least one Non Line of Sight (NLOS).
\mbox{Figure \ref{fig:testbed}} visualizes the positions with markers in green (LOS) and red (NLOS).
\\However, before the localization accuracy at the individual positions can be evaluated, another factor must first be considered. 
Until now, this article solely presented the process of estimating individual TDoA values, necessary for performing the localization.
No consideration was given to the actual multilateration algorithms, which allow for 2D position estimation based on the measured TDoA values.
However, the solution to this multilateration problem is already well described in the literature.
Thus, this article follows the considerations from \cite{TDoA_2}.
The authors highlighted that the iterative Foy multilateration solver shows best accuracy.
However, the algorithm requires an initial position estimate for convergence.
Thus, a second solver is executed prior to the precise Foy solver.
This allows to obtain an initial estimate for the more precise Foy solver.
In the prevailing case, this is achieved by using the Fang multilateration solver.
This combination is expected to operate near the optimum.
However, future work might include more detailed investigations on how close the method operates near the multilateration CRB.
\\Coming back to the evaluation of the localization accuracy, the now stated multilateration method closes the gap between measured TDoA values and position estimates.
In the following, all endpoint positions from \mbox{Figure \ref{fig:testbed}} are evaluated and compared.
For every position, at least 50 measurements are taken to achieve a statistical relevance of the measured localization accuracy.
Single outliers are excluded from the measurement.
This concerns deviations of the single TDoA value by more than 1000\,m from the median of the respective position.
However, the number of outliers never exceeded 10 \, \% of the total number of measurements for each position tested.
The ground truth of the position information during the measurements was obtained by a Columbus P10\footnote{https://www.columbus-gps.de/produkte/columbus-p10-pro-gnss-datenlogger, accessed March 2026} GNSS data logger.
Finally, \mbox{Figure \ref{fig:meas_ilm_nlos_combined}} shows the cumulative density function (CDF) for the localization error.
In detail, \mbox{Figure \ref{fig:meas_ilm_nlos_zoomed}} visualizes the comparison of the CDFs for the endpoints with LOS and NLOS condition.
The CDF for the NLOS case is zoomed to the 150\,m.
In addition, \mbox{Figure \ref{fig:meas_ilm_nlos}} shows the complete CDF for the NLOS case.
\\The considerations start with the observations from \mbox{Figure \ref{fig:meas_ilm_nlos_zoomed}}.
The position estimates for the LOS condition are very accurate.
In detail, the 50\,\% error percentile ($P_{50}$) is just below 10\,m, whereas the 90\,\% error percentile ($P_{90}$) ranges at approximately 17\,m.
This underscores the functionality of the system presented here. 
Furthermore, it demonstrates the potential for high-precision GNSS-free localization.

However, the required LOS propagation is a restrictive requirement that is rarely met in urban areas.
Thus, the LOS condition is directly compared to the NLOS condition.
At least one blocked LOS is a far more realistic constraint on the localization in dense urban environments.

Coming back to the observations in \mbox{Figure \ref{fig:meas_ilm_nlos_zoomed}}, these suboptimal conditions manifest in an increased localization error.
This can be traced back to multipath propagation within the channel between the endpoint and the individual base stations.
This causes an erroneous estimate of the TDoA value, and thus, an erroneous position estimate. 
The multipath influence causes an increase of the $P_{50}$ to approximately 40.2\,m.
This suggests that the presented localization approach still yields sufficiently good results even under multipath conditions.

However, a more interesting and potential harmful observation concerns the 90\,\% error percentile ($P_{90}$), visible in \mbox{Figure \ref{fig:meas_ilm_nlos}}.
It ranges at approximately 430\,m, causing significant errors in some measurements.
Again, multipath propagation is the explanation.
One might think that such long multipaths, presumably caused by reflections off mountains near the measurement area could be easily distinguished from the LOS path.
To understand this effect, we need to take another look at the measurement method.
The TDoA estimation is based on a simple peak finding algorithm, as described in \eqref{eq:tdoa_extraction}.
However, this method is critical if the LOS has a lower amplitude than the multipath.
In these cases, the wrong hypothesis is considered as the TDoA value, causing significant error tail in the localization.
Future work might include approaches to solve this issue.
This includes the usage of thresholding, where not only the most prominent peak within the CCF is used but also peaks with a lower amplitude.
Based on that, multiple hypothesis might be checked for plausability.
Further approaches include the usage of different estimation methods, such as Multiple Signal Classification (MUSIC) to estimate the TDoA values.
Ultimately, the choice of the reference station and the exclusion of base stations subject to multipath interference from the localization process could also help mitigate these effects.

    \section{Summary and Conclusion}
\label{Conclusion}
This article presented an approach for the Global Navigation Satellite System (GNSS)-free localization of endpoints within the Internet of Things (IoT).
The presented approach relies on the extraction of Time Difference of Arrival (TDoA) values
from the communication waveform.
In the prevailing case, the considerations were carried out for the Low Power Wide Area Network (LPWAN) standard according to ETSI TS 103 357 with brand name mioty.

In detail, the article contained a concept for precise extraction of the TDoA values from the communication waveform, taking into account the incoherent frequency synthesis of the Frequency Hopping (FH) signal structure by using a differential correlation approach between multiple base stations.
Furthermore, the article made use of a highly precise time and frequency synchronization approach, relying on the usage of broadcast signals as so-called Signals of Opportunity (SoO).

Based on the stated concept, the theoretical limits according to the Modified Cramer Rao Lower Bound (MCRB) were derived, promising sub-meter localization accuracy for even low signal levels.
A waveform modification was presented to further enhance the localization accuracy.
Based on the concept, simulations were carried out to highlight the influence of the Telegram Splitting Multiple Access (TSMA) signal structure on the localization accuracy.
Further, simulations have proven the functionality of the concept under the influence of Additive White Gaussian Noise (AWGN) and residual frequency offsets.
A final step included measurements in a cable-based environment, achieving standard deviations down to 0.3\,m.

Finally, an $9\,\text{km}^2$ wide installation in an urban area showed that the stated concept allows a 2D localization error with a 50\,\% percentile ($P_{50}$) below 10\,m and a $P_{90}$ below 17\,m under pure Line of Sight (LOS) conditions.
With Non Line of Sight (NLOS) conditions, the $P_{50}$ increased to 40.2\,m, whereas a relatively low number of high outliers increased the $P_{90}$ to 430\,m.
Despite this increased error tail in NLOS scenarios, the presented TDoA localization principle is a very promising candidate for accurate positioning.
Similar measurements were also conducted with established testbeds in more rural environments, achieving accuracy in the range of the LOS scenario.

Future work might investigate the issue with the high outliers by using more sophisticated multipath resolution methods.
Additionally, a further increase of the usable bandwidth might relax the multipath problem.
Another issue that needs further improvement is the current necessity of IQ data exchange between the base stations in order to perform the differential correlation and synchronization.
However, current work already allows to reduce the constraints on the network capability from multiple tens of MBit/s to kBit/s for the synchronization data \cite{SynchSebastianConceptCows, SynchSebastianMinimizedTraffic, SynchSebastianCompression}.
Similiar modifications might be applied to the LPWAN data.
\section*{Acknowledgment}
This work is part of the research project 5G-Flexi-Cell (grant no. 01MC22004B) funded by the German Federal Ministry for Economic Affairs and Climate Action (BMWK) based on a decision taken by the German Bundestag.
    
    \appendices
\section{Derivation of the MCRB for TDoA Estimation}\label{chap:App_MCRB_Derivation}

The derivation of MCRB($\tau$) starts by using the signal model of a GMSK-based waveform, formerly described in \mbox{Chapter \ref{chap:Concept}}.
However, the random phase offsets at the individual bursts are discarded to obtain the estimate of the achievable estimation accuracy using the full bandwidth.
This reflects the case of fully coherent correlation with the prototype function, as discussed in an earlier part of this paper. 
This assumption is justified, as \mbox{Chapter \ref{chap:Concept}} has already shown that the differential correlation leads to an extinction of these offsets.
After discarding random phase offsets from \eqref{eq:ep_dig}, the model for a signal radiated from the endpoint can be stated as:
\begin{IEEEeqnarray}{rCl}
s_{EP}(t)
& = & \sum_{b = 0}^{B-1} A
      \exp\Big(-j(2\pi\epsilon t + \phi_c)\Big) \nonumber \\
&   & \cdot \exp\Big(
      j2\pi \sum_{k=0}^{K-1}
      (a[k] + 4T f_B[b]) \nonumber \\
&   & \qquad \cdot
      g(t - kT(1+\xi) - \tau)
      \Big)
\end{IEEEeqnarray}
The attentive reader might notice an assumption in the signal model.
The single bursts are not shifted in the time domain, as it is actually the case for the TSMA waveform.
The signal model in the case at hand assumes that all $B$ bursts are simultaneously active.
One possibility is to model the more complex TSMA structure by multiplication with boxcar functions for the individual bursts.
However, this approach is not recommended, as it would result in the assumption of infinitely steep power ramps of the transceiver.
In the frequency domain, this results in an unrealistically large bandwidth of the transmit signal, resulting in an overestimated estimation accuracy.
Therefore, the simplified model is used below.

Coming back to the derivation of the MCRB, \cite{MCRB} states the starting point for the derivation of the MCRB.
It can be given as:
\begin{equation}
    \text{MCRB}(\lambda) = \frac{N_0}{E_u\bigg\{\int_{T_0}\Bigl|\frac{\partial s(t)}{\partial \lambda}\Bigl|^2\bigg\}},
\label{eq:MCRB}
\end{equation}
showing the necessity to calculate the derivative of the waveform with respect $\lambda$ with a subsequent formation of the expected value over $\bm{u}$.
In the prevailing case, the MCRB has to be derived for the timing information of the waveform.
Thus, the relation $\lambda = \tau$ holds.
The interfering parameters are summarized in the nuisance parameter vector \mbox{$\bm{u} \in \{ \phi, \epsilon, \xi, \bm{a}, \bm{f_B}\}$}.
This vector contains the carrier phase offset $\phi_c$, the CFO $\epsilon$, the SCO $\xi$, and the modulated data symbols $\bm{a}$ with running index k.
Further, it depends on the carrier frequencies $f_B[b]$ with running index $b$.
Further, the stated signal model matches a CPM waveform with an amplitude of $A=\sqrt{\frac{2E_S}{T}}$.
To match the CPM waveform, a GSMK modulation, the pulse shape $g(t)$ has to be modified.
According to \cite{Proakis2001}, it follows:
\begin{equation}
    g(t)=\begin{cases}
			0, & \text{if $t$ $\leq$ 0}\\
            \frac{t}{4T}, & \text{0 $\leq$ $t$  $\leq$ T} \\
            \frac{1}{4}, & \text{$t$ $\geq$ T},
		 \end{cases}
	\label{eq:g_t}
\end{equation}
The consideration begins by deriving the transmitted endpoint signal with respect to the parameter $\tau$.
Using the exponential derivative rule, it follows:
\begin{IEEEeqnarray}{rCl}
\frac{d s_{EP}(t)}{d \tau}
& = &
-\sum_{b=0}^{B-1} A
\exp\!\Big(
-j(2\pi\epsilon t + \phi_c)\Big)
\nonumber\\
& & \quad
\cdot \Big( j2\pi
\sum_{k=0}^{K-1}
\big(a[k] + 4T f_B[b]\big)
\nonumber\\
& & \qquad \cdot
g\!\left(t-kT(1+\xi)-\tau\right)
\Big)
\nonumber\\
& & \cdot j2\pi
\sum_{k=0}^{K-1}
\big(a[k] + 4T f_B[b]\big)
\,\psi_k(t,\tau)
\label{eq:sep_derivative}
\end{IEEEeqnarray}
With $\psi_k(t,\tau)$ being a helper function defined as:
\begin{equation}
\psi_k(t,\tau) \triangleq \frac{d}{d\tau}
g\!\left(t-kT(1+\xi)-\tau\right)
\label{eq:psi_def}
\end{equation}
After the derivation of the exponential, the equation breaks down into two parts, namely the unchanged exponential part and the derived part of the exponential. 
Since \eqref{eq:MCRB} uses only the absolute value of the derivation, the exponential part can be dropped, yielding:
\begin{equation}
\left|
\frac{ds_{EP}(t)}{d\tau}
\right|^2
=
4\pi^2 A^2
\left|
\sum_{b=0}^{B-1}\sum_{k=0}^{K-1}
\left(a[k] + 4T f_B[b]\right)
\psi_k(t,\tau)
\right|^2 
\label{eq:dep}
\end{equation}
This is particularly interesting since the MCRB no longer depends on the nuisance parameters $\phi$ and $\epsilon$. 
However, it still depends on the encoded data symbol vector $\bm{a}$ and the carrier frequency vector of the individual bursts $\bm{f_B}$.
In the next step, the expectation value over the remaining nuisance parameters $\bm{u} \in \{ \bm{a}, \bm{f_B}\}$ has to be calculated according to \eqref{eq:MCRB}.
After applying a row of simple modifications, it follows:
\begin{equation}
\begin{aligned}
E_{\bm{u}}\!\left\{\left|\frac{d s(t)}{d\lambda}\right|^2\right\}
= E_{\bm{a},\bm{f_B}}
\!\left\{\left|\frac{d s_{EP}(t)}{d\tau}\right|^2\right\}
\\[4pt]
\begin{aligned}
&= 4\pi^2 A^2 E_{\bm{a},\bm{f_B}}\!\Bigl\{
\sum_{b=0}^{B-1}\sum_{k=0}^{K-1}
\bigl|\bigl(a[k]+4T f_B[b]\bigr)\psi_k(t,\tau)\bigr|^2
\Bigr\}
\\
&= 4\pi^2 A^2 E_{\bm{a},\bm{f_B}}\!\Bigl\{
\sum_{b=0}^{B-1}\sum_{k=0}^{K-1}
\bigl(a^2[k]+8T f_B[b]a[k]+16T^2 f_B^2[b]\bigr)
\\
&\qquad\qquad\cdot \bigl|\psi_k(t,\tau)\bigr|^2
\Bigr\}
\end{aligned}
\end{aligned}
\label{eq:sep_to_tau_sq}
\end{equation}
However, the transformation is only valid because at any given time at most one symbol-derivative contribution is nonzero, so all cross-terms between different symbol indices vanish as well as and cross-terms between different bursts since the bursts occupy non-overlapping carrier frequencies.
Furthermore, remembering the GMSK modulation, any data symbol is selected from the alphabet \mbox{$a[k] \in \{-1, 1\}$}.
Without loss of generality, it directly follows: 
\begin{equation}
    E_{\bm{a},\bm{f_B}}\{a^2[k]\} = 1
\end{equation}
Furthermore, taking into account the possible value range of \mbox{$a[k] \in \{-1, 1\}$}, and hence $E_{\bm{a}}\{\bm{a}\} = 0$, and its statistical independency from $\bm{f_B}$, it follows:
\begin{equation}
\begin{aligned}
E_{a,f_B}\!\left\{8T f_B[b]\,a[k]\psi_k(t,\tau)\right\}
\end{aligned}
\end{equation}
\begin{equation}
\begin{aligned}
&= 8T\psi_k(t,\tau)\,
   E\!\left\{f_B[b]\right\}E\!\left\{a[k]\right\} \\
&= 8T\psi_k(t,\tau)\cdot 0 \\
&= 0 .
\end{aligned}
\label{eq:expectation_zero}
\end{equation}
At this point, \eqref{eq:sep_to_tau_sq} boils down to:
\begin{equation}
\begin{gathered}
E_{\bm{a},\bm{f_B}}\!\left\{
\left|\frac{d s_{EP}(t)}{d \tau}\right|^2
\right\}
= \\[2pt]
4\pi^2 A^2
\sum_{b=0}^{B-1}\sum_{k=0}^{K-1}
E_{\bm{a},\bm{f_B}}\!\Bigg\{
\left(1+16T^2 f_B^2[b]\right)
\left|\psi_k(t,\tau)\right|^2
\Bigg\}
\end{gathered}
\label{eq:E_a_b_simplified}
\end{equation}
By calculating the derivation for \eqref{eq:g_t}, it follows: 
\begin{equation}
    \bigl|\psi_k(t,\tau)|^2=\begin{cases}
            \frac{1}{16 T^2}, & \text{if $0 \leq t \leq T$}\\
            0, & \text{otherwise} 
            \end{cases}  
\label{eq:g_t_2}
\end{equation}
Furthermore, the sum $\sum_k$ from \eqref{eq:E_a_b_simplified} can be dropped as \eqref{eq:g_t_2} shows that the derivative of the pulse shape is only specified within $0 \leq t \leq T$.
It follows:
\begin{equation}
	E_{\bm{a},\bm{f_B}}\bigg\{\Bigl|\frac{d s_{EP}(t)}{d \tau}\Bigl|^2\bigg\} = \frac{4\pi^2 A^2}{16\,T^2} \Big [B + 16T^2 \sum_{b=0}^{B-1} E_{\bm{f_B}}\Big \{f_B^2[b]\Big\} \Big ]
	\label{eq:E_a_b_simplified_II}
\end{equation}
Using the carrier frequency of each burst directly is of little use, as the MCRB would depend on the frequency band in which the transmission takes place.
Therefore, the following considerations aim to find a starting point for this expression.
The expression is only independent of the absolute carrier frequencies when the single carrier of the frequency hopping waveform is centered around the origin $f=0\,$Hz, or in other words, the expectation value over all carrier frequencies is 0.
The expression can be stated as:
\begin{equation}
		E_{\bm{f_B}} \Big\{f_B^2[b] \Big\} =  \text{Var}\Big \{ f_B[b]\Big \} + \underbrace{E^2 \Big\{ f_B[b]\Big\}}_{=0} = \sigma^2_f
\end{equation}
After using this finding in \eqref{eq:E_a_b_simplified_II}, it directly follows:
\begin{IEEEeqnarray}{rCl}
E_{\bm a,\bm f_B}\!\Bigg\{
\Bigl|\frac{d s_{EP}(t)}{d\tau}\Bigr|^2
\Bigg\}
& = &
\frac{4\pi^2 A^2}{16\,T^2}
\Big[
B + 16T^2
\sum_{b=0}^{B-1} \sigma_f^2
\Big]
\IEEEnonumber\\[4pt]
& = &
\frac{4\pi^2 A^2}{16\,T^2}
\big(1 + 16T^2 \sigma_f^2\big)\,B 
\label{eq:E_a_b_simplified_III}
\end{IEEEeqnarray}
According to \eqref{eq:MCRB}, the final step now involves the integration over the whole observation length $T_0$. 
Since there is no dependency on $t$, the integration boils down to a simple multiplication.
Furthermore, the observation length equals exactly the duration of a single telegram.
The signal model already assumes that $B$ bursts are active in parallel.
Hence, to match the energy of a whole telegram, the observation length equals exactly the duration of one burst.
Assuming one burst consists of $L$ symbols, it follows that $T_0 = L\cdot T$, with $T$ being the symbol rate.
It finally follows:
\begin{IEEEeqnarray}{rCl}
E_{\bm u}\!\Bigg\{
\int_{T_0}
\Bigl|\frac{d s_{\text{EP}}(t)}{d\lambda}\Bigr|^2
\Bigg\}
& = &
\int_{LT}
E_{\bm a,\bm f_B}\!\Bigg\{
\Bigl|\frac{d s_{\text{EP}}(t)}{d\tau}\Bigr|^2
\Bigg\} dt
\IEEEnonumber\\[4pt]
& = &
\frac{\pi^2 L \,B\,E_S
\big(16T^2\sigma_f^2 + 1\big)}
{2T^2}.
\label{eq:mcrb_nearly}
\end{IEEEeqnarray}
And by putting \eqref{eq:mcrb_nearly} into \eqref{eq:MCRB} again, the MCRB can be stated as a closed-form solution:
\begin{equation}
    \text{MCRB}(\tau) = \frac{2T^2}{\pi^2 L \,B\,(16 T^2 \sigma_f^2 + 1)}\frac{N_0}{E_S}
\label{MCRB_CLOSED}
\end{equation}

\section{Considerations on the mioty Waveform Parameters according to ETSI TS 103 357}\label{chap:App_mioty_param}

\begin{figure}[t]
\centerline{\includegraphics[width=0.5\textwidth]{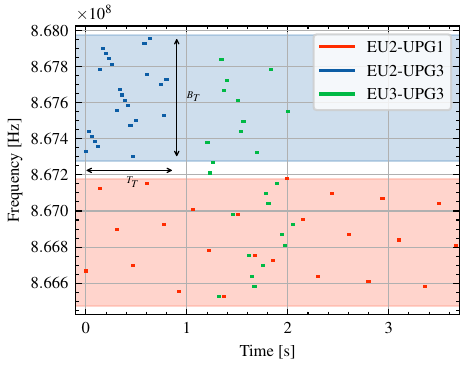}}
\caption{Time and frequency representation of the TSMA-based mioty waveform in the 868\,MHz band. The blue area denotes the upper and the red area of the lower frequency band.}
\label{fig:mioty_pattern}
\end{figure}
This section investigates the mioty waveform, as specified in \cite{ETSI}, and its relevant parameters for localization in detail.
The mioty waveform uses the so-called Telegram Splitting Multiple Access (TSMA) technology.
This principle drastically increases the interference robustness and channel capacity \cite{mioty_0}.
A single packet, called telegram, is split into multiple subpackets, called bursts. 
In general, one mioty telegram can carry a payload of 10 Bytes, consisting of a total of 24 bursts.
For bigger payloads, the telegram can be extended by so-called extension bursts.
Each burst has a duration of approx. 15.1\,ms, totalling 36 symbols modulated at a symbol rate of 2.38\,kHz.
The modulation itself uses Gaussian Minimum Shift keying (GMSK) with a bandwidth ratio of \mbox{BT = 1}.

According to \cite{ETSI}, the mioty waveform further offers different transmission modes that differ in the occupied bandwidth $(B_T)$ as well as the transmission duration $(T_T)$.
A transmission with narrow bandwidth is assigned to the bandwidth mode EU1.
It has a total bandwidth of roughly 60.2\,kHz.
A parameter denoted as carrier spacing (CS) represents the spacing between two carriers in the frequency domain.
Assuming a total of 24 bursts, the EU1 bandwidth mode features a CS of 2.38\,kHz.
In case of the more broadband bandwidth mode EU2, the CS is increased by a factor of 12, which results in a total bandwidth of roughly 688.6\,kHz with a CS of 28.56\,kHz.
Another distinguishing feature concerns the transmission duration of individual telegrams.
The standard transmission duration, denoted as Uplink Pattern Group (UPG) 1, has a total duration of approximately 3.6\,s.
This is achieved by adding temporal pauses between the transmission of single bursts.
Within the UPG1, a total of 8 different patterns is defined, enhancing the collision robustness in case two endpoints start their transmission at once.
The transmitting endpoint randomly chooses its pattern from the given table of possible patterns.
These patterns differ in the sequence of frequency slots used and in the varying pause times.
However, besides UPG1, there exists also the UPG3 group.
It is designed for ultra low latency with a total transmission duration of 0.86\,s.
This is achieved by shortening the pauses between consecutive bursts.
\mbox{Table \ref{tab:eu_params}} gives an overview over the different specified bandwidth modes and UPG modes.
\begin{table}[t]
\caption{Transmission parameters for mioty according to \cite{ETSI}.}
\centering
\small
\begin{tabular}{|c||c|c||c|c||c|}
\hline
\multirow{2}{*}{} 
    & \multicolumn{2}{c||}{EU1} 
    & \multicolumn{2}{c||}{EU2} 
    & EU3 \\ \cline{2-6}
 & UPG1 & UPG3 & UPG1 & UPG3 & UPG3 \\ 
\hline
$T_{T}$ [s] 
    & 3.6 & 0.86 
    & 3.6 & 0.86 
    & 0.86 \\ 
\hline
$B_{T}$ [kHz] 
    & 60.2 & 60.2 
    & 688.6 & 688.6 
    & 1310.2 \\ 
\hline
CS [kHz] 
    & 2.3 & 2.3 
    & 28.5 & 28.5 
    & 57.1 \\ 
\hline
\end{tabular}
\label{tab:eu_params}
\end{table}

Another aspect becomes apparent when inspecting the individual EU/UPG modes, as shown in \mbox{Figure \ref{fig:mioty_pattern}}.
For reasons of simplicity, only the wider bandwidth mode EU2 is displayed.
The whole frequency band is divided in the so-called upper frequency band (depicted in blue) and the lower frequency band (depicted in red).
The endpoint randomly chooses one of the two frequency bands for one transmission.

However, this principle also leaves room for improvement.
By using both channels in combination during a single transmission, the effective bandwidth could be doubled in theory.
This should also benefit runtime-based localization, where bandwidth is one of the most important tuning parameters.
To this end, a new pattern was designed that uses the doubled carrier spacing (CS = 57.12\,kHz) in comparison to the standard EU2.
The number of bursts remains constant at 24 and the pauses between the individual bursts are fixed to the UPG3 pattern.
By doing so, the effective bandwidth $B_T$ can be enhanced to approximately 1.31\,MHz.
A further advantage of this method is the full compatibility with the existing decoder architecture.
In the current implementation, the decoder listens to both frequency bands (upper and lower).
After detection of a preamble, the decoder checks if the found bursts correspond to an existing pattern in the lookup table.
By simple adding the newly postulated pattern to this lookup table, the decoder is capable of detecting the newly postulated pattern, in future denoted as bandwidth mode EU3.
%

\begin{IEEEbiography}[{\includegraphics[width=1in,height=1.25in,clip,keepaspectratio]{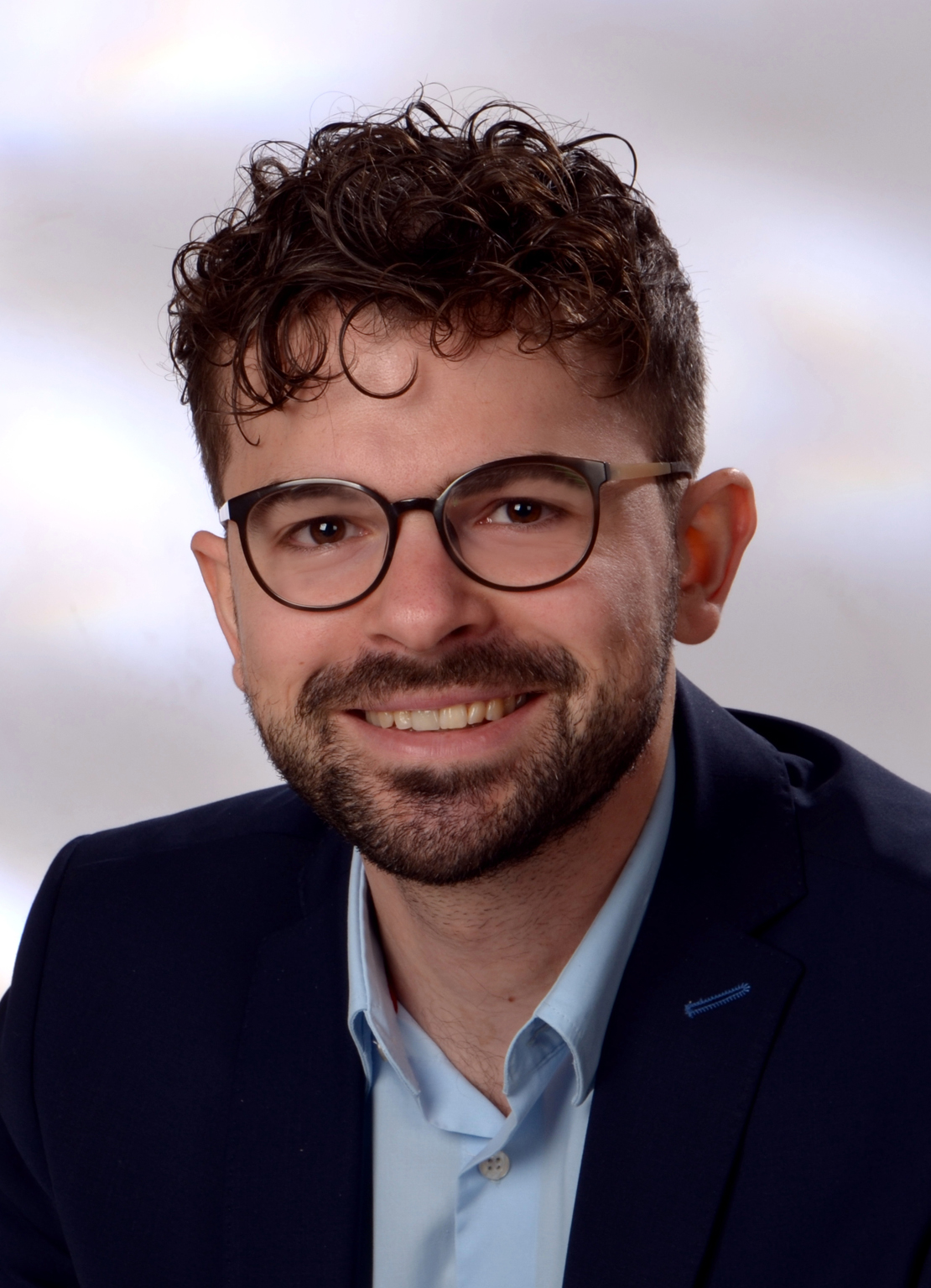}}]{Thomas Maul} 
received the B.Sc and M.Sc degrees in electrical engineering and information technology from Friedrich-Alexander-University Erlangen-Nuremberg, Erlangen, Germany, in 2019 and 2021, respectively. From 2021 to 2025 he was part of the Dependable Machine-to-Machine Communication research group at TU Ilmenau, where he pursued his Dr.-Ing degree. 
\\He is a Researcher with the Fraunhofer Institute of Integrated Circuits IIS, Nuremberg. His research interests include the development of algorithms for runtime-based localization in LPWAN (Low Power Wide Area Networks) and synchronization using Signals of Opportunity.
\end{IEEEbiography}

\begin{IEEEbiography}[{\includegraphics[width=1in,height=1.25in,clip,keepaspectratio]{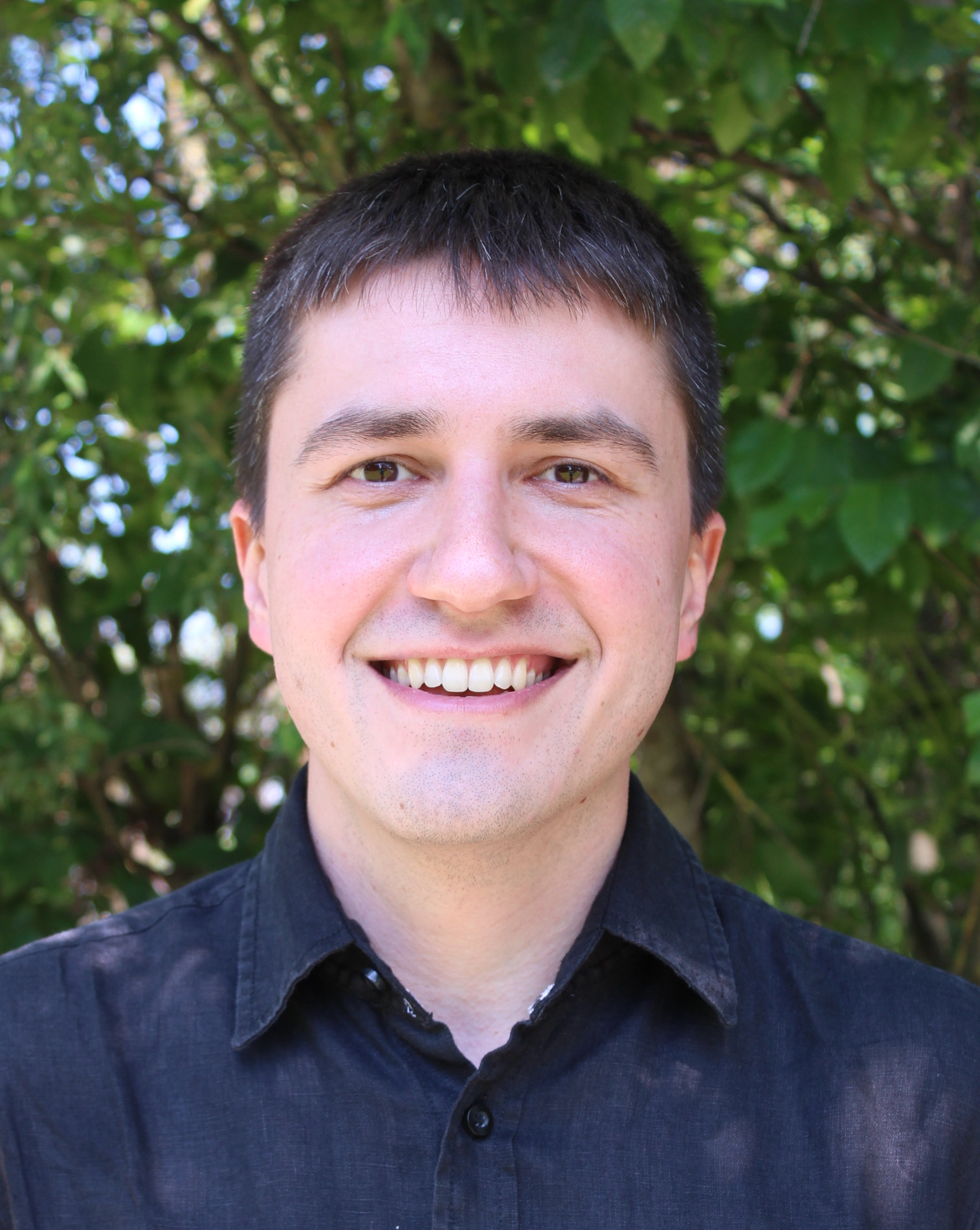}}]{Alfred Mueller}
obtained B.Sc. degrees in Information Technology and Computer Science from the University of Klagenfurt, Austria in 2015 and 2016. He then continued his studies in Electrical Engineering at the Friedrich-Alexander-University of Erlangen-Nuremberg, Germany, receiving an M.Sc. degree in 2018.
\\As a researcher at the Fraunhofer Institute for Integrated Circuits in Nuremberg from 2018 to 2020, he focused on the inductive near-field localization system IndLoc as well as localization approaches for the Low Power Wide Area Network (LPWAN) technology Mioty.
\end{IEEEbiography}

\begin{IEEEbiography}[{\includegraphics[width=1in,height=1.25in,clip,keepaspectratio]{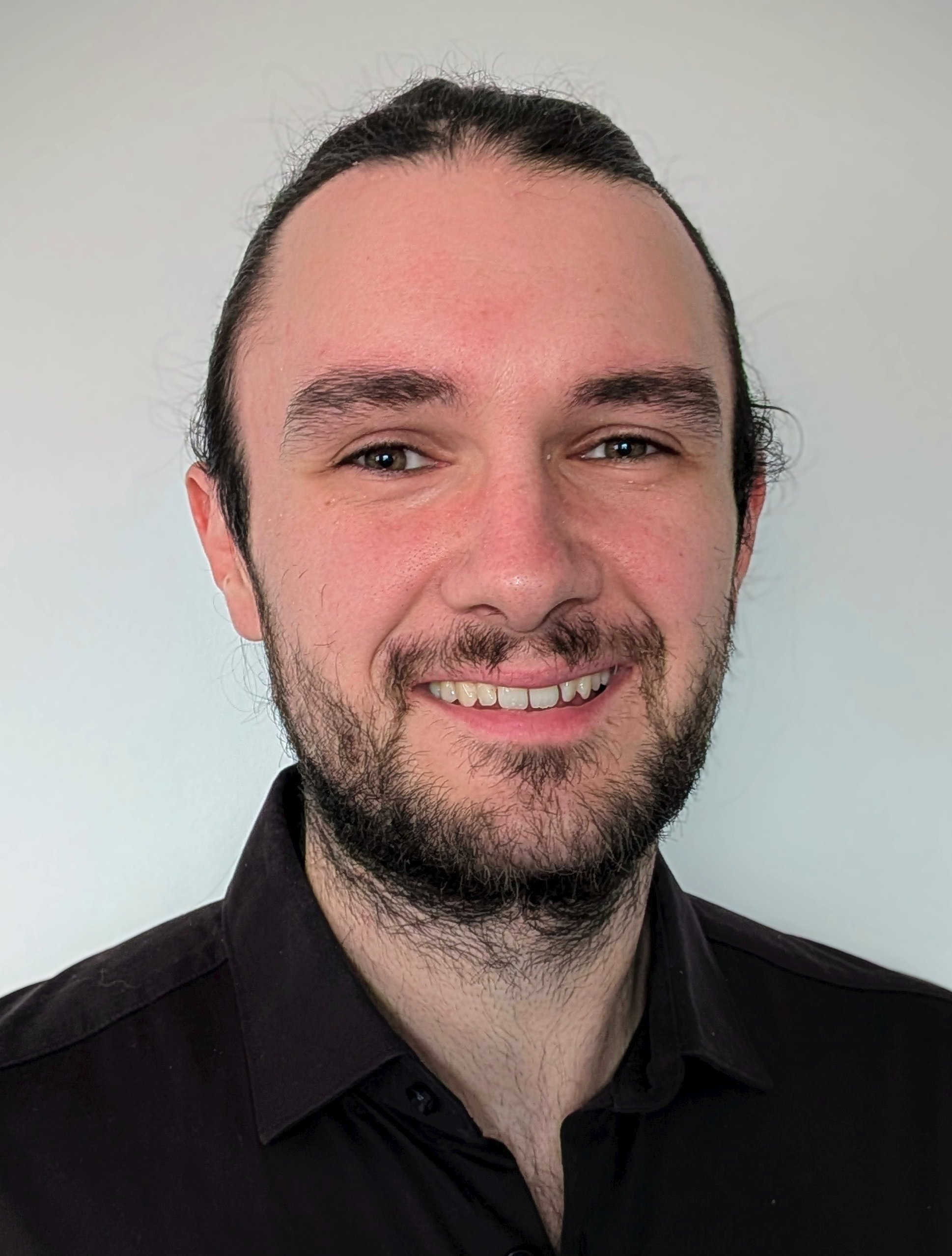}}]{Sebastian Klob}
received the B.Sc. and M.Sc. degrees in Electrical Engineering from Friedrich-Alexander University (FAU) Erlangen-Nuremberg in 2019 and 2021, respectively. Afterwards, he started a doctorate at the LIKE Chair at Friedrich-Alexander-Universität Erlangen-Nürnberg, where he conducts research on ultra-efficient and ultra-precise synchronization with Signals of Opportunity. 
\end{IEEEbiography}


\begin{IEEEbiography}[{\includegraphics[width=1in,height=1.25in,clip,keepaspectratio]{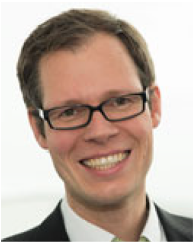}}]{Joerg Robert}
 studied Electrical Engineering and Information Technology at TU Ilmenau and TU Braunschweig, Germany. 
 From 2006 to 2012, he conducted research on digital television at the Institute of Communications Engineering at TU Braunschweig, where he played a key role in the development of DVB-T2, the second generation of digital terrestrial television. 
 In 2013, he completed his PhD at TU Braunschweig on the topic of Terrestrial TV Broadcast using Multi-Antenna Systems. 
 In 2012, he joined the LIKE Chair at Friedrich-Alexander-Universität Erlangen-Nürnberg (FAU), Germany, where he pursued research on Low Power Wide Area Networks (LPWAN). 
 From 2021 to 2024, he served as a full professor and head of the Group for Dependable Machineto-Machine Communication at the Department of Electrical Engineering and Information Technology at Technische Universität Ilmenau, Germany. 
 Since 2025, he is a full professor specializing in localization systems at the LIKE Chair at Friedrich-Alexander-Universität Erlangen-Nürnberg. 
 Additionally, he is affiliated with the Fraunhofer Institute for Integrated Circuits (IIS) in Nürnberg. 
 
 Joerg Robert is actively involved in international standardization efforts. 
 He currently serves as the secretary of the IEEE 802.15 standardization group and vice-chair of several wireless standardization groups within IEEE 802.
\end{IEEEbiography}




\end{document}